\documentclass[12pt]{article}
\input epsf.sty
\newcommand{\nl}{\hspace{-.65cm}}
\newcommand{\be}{\begin{equation}}
\newcommand{\ee}{\end{equation}}
\newcommand{\ben}{\begin{eqnarray}\displaystyle}
\newcommand{\een}{\end{eqnarray}}

\newcommand{\p}{\partial}

\def\sqr#1#2{{\vcenter{\vbox{\hrule height.#2pt
         \hbox{\vrule width.#2pt height#1pt \kern#1pt
            \vrule width.#2pt}
         \hrule height.#2pt}}}}

\begin{document}

%here

{}~ \hfill\vbox{\hbox{hep-th/0604190} \hbox{PUPT-2196} }\break

\vskip 1cm

\begin{center}
{\bf A COMMENT ON TECHNICAL NATURALNESS AND\\ THE COSMOLOGICAL
CONSTANT }

\vspace{10mm}

\normalsize{Nissan Itzhaki }

\vspace{10mm}

\normalsize{\em Physics Department, Princeton University,
Princeton, NJ 08544}\end{center}\vspace{10mm}

\begin{abstract}

\medskip

We propose a  model of dynamical relaxation of the cosmological
constant. Technical naturalness of the model and the present value
of the vacuum energy density imply an upper bound on  the
supersymmetry breaking scale and the reheating temperature   at
the  TeV scale.

\end{abstract}

\newpage

\baselineskip=18pt

%\tableofcontents

\newpage

From an effective  field theory perspective the problem with the
cosmological constant term is that it is a relevant term that
seems to be small for no apparent reason. In supersymmetric
theories the cosmological constant is of the order of $M_{\rm
SUSY}^4$ (where $M_{\rm SUSY}$ is the
 supersymmetry breaking scale) which is at least $10^{-60}$ in
 Planck units,
  while the actual value of the vacuum
energy density is only about $ 10^{-120}$. The growing evidence
for inflation (for recent results that support  inflation see
\cite{wmap3}) makes the problem  even more intriguing since
inflation is driven by vacuum energy much larger than $10^{-120}$.
Accordingly, the natural question is: why is the ratio of the
vacuum energy during inflation to the current vacuum energy so
large (and yet not infinite)?

At least conceptually the simplest resolution  to this question is
that there is a dynamical adjustment mechanism that ensures the
smallness of the vacuum energy at the end of inflation. Assuming
such a mechanism can be described by a low energy effective action
the challenge is to come up with  an effective field theory that
appropriately adjusts the vacuum energy and is natural, or at
least technically natural. More than twenty years ago Abbott
proposed a technically natural model that does exactly that
\cite{ab}. However, as pointed out by Abbott himself, the problem
with his model is that the vacuum energy is adjusted via
tunnelling from one local minimum  to the next. Hence, much like
in old inflation \cite{oi}, the resulting universe is empty,
containing neither radiation nor matter.

In this note we attempt  to improve on Abbott's model by proposing a
model that evades   the emptiness problem and appears to be
technically natural.\footnote{A different approach to the emptiness
problem was suggested recently in \cite{pa}.} The effective action
we propose takes the form (we work in units where $M_{Pl} =(8\pi
G)^{-1/2}=1$)
 \be S= S_{EH}~+~ S_{\rm relaxation  }~+~S_{\rm inflation}.\ee
$S_{EH}$ is the  Einstein-Hilbert action and  the relaxation
action is
 \be\label{sl} S_{\rm relaxation }=\int d^4 x\sqrt{-g} \left(
 %+ {R \over 2}
 -{1 \over 2} (\p \psi)^2 - V_{\rm ren} -V(\psi)
 \right). \ee
$V_{\rm ren}$ is the renormalized value of the cosmological constant
including radiative contributions from all fields.  We assume that
$V_{\rm ren}$ is large and positive. The role of $S_{\rm
relaxation}$ is to reduce the vacuum energy slowly in a controlled
fashion. For this reason we follow \cite{ab} (see also
\cite{Banks:1984tw,Banks:1984cw,Linde:1986dq})  and take
 $$V(\psi)=\epsilon\:\psi ~~~~~ \mbox{where} ~~~~~ | \epsilon | \ll 1.$$
 The condition
$ | \epsilon | \ll 1$  is essential for the model and must not
receive large quantum corrections. Since $\epsilon \: \psi $ is a
relevant term one might suspect that small $\epsilon$ is as
unnatural as a small cosmological constant. This, however, is not
the case since $\epsilon\phi$, unlike the cosmological constant,
is protected by a symmetry: when $\epsilon=0$ the action is
invariant under the shift symmetry $\psi\to \psi +const.$ which
ensures that corrections to $\epsilon$ are proportional to
$\epsilon$. Namely $ | \epsilon | \ll 1$ is technically natural.

As a result of this small slope $\psi$ will roll down the
potential reducing the vacuum energy. The slow roll approximation
is valid as long as the vacuum energy is much larger than
$\epsilon$. As argued above we can take $\epsilon$ to be
arbitrarily small so the slow roll approximation is valid at
arbitrarily small vacuum energy. This way we are able to reduce
the vacuum energy but not to  convert it into kinetic energy,
which is necessary to avoid the emptiness problem.

The role of $ S_{\rm inflation}$ is to fix that  while making sure
that the vacuum energy at the end of inflation is small. There are
various actions one might wish to consider. Below we study the one
that we believe is the simplest single-field $ S_{\rm inflation}$
that illustrates how the adjustment mechanism works. We do not
address phenomenological aspects of the model
%(such as COBE normalization)
nor possible realization in string theory. These
issues are likely to require a more complicated multi-field
action. The action takes the form
 \be\label{wf} S_{EH}+ S_{\rm inflation}= \int d^4 x \sqrt{-g}
  \left[ \frac12 e^{- \phi^2} R
   -V(\phi)
\right]+ \int d^4 x {\cal L}_{kin}. \ee The first term is familiar
from supergravity theories before rescaling to the Einstein frame.
${\cal L}_{kin}$ is a bit complicated in this frame and is fixed
by  requiring that in the Einstein frame it takes the standard
form (see (\ref{ei})). The details of ${\cal L}_{kin}$ (that can
be found for example in chapter 21 of \cite{wb}) will play no role
below. $V(\phi)$ is designed to have the following properties. It
has a maximum
  at $\phi=0$
$$ \gamma \equiv - \left. \frac{d^2V(\phi)}{d\phi^2} \right| _{\phi=0} > 0 ,$$
and the difference between  its values at the maximum and the
minimum is
 \be\label{con} \Delta V \equiv V_{max}-V_{min} = \frac{\gamma}{4} .\ee

$S_{\rm inflation}$ modifies  the dynamics  in the following way.
Expanding the first term in (\ref{wf}) we see that the effective
mass of $\phi$  is $m_{\rm eff}^2= R-\gamma$. In the slow roll
approximation the relationship between the curvature and the vacuum
energy density is $ R= 4 V$. Therefore,  for $V> \gamma/4$ we have
$m_{\rm eff}^2>0$  and $\phi$ does not receive an expectation value
and does not affect  the FRW equations that are
 controlled by $\psi$. When the vacuum energy density crosses the
critical value $V_c=\gamma/4$ an instability is developed and $\phi$
acquires an expectation value. This reduces $R$ which further
reduces $m_{\rm eff}^2$ which  in turn increases the expectation
value of $\phi$ even further. Eq.(\ref{con}) ensures that the end
result of this feedback mechanism is a space with no vacuum energy
at all. This follows from the fact that for $R=0$ the effective
potential for $\phi$ is simply $V(\phi)$ and so the vacuum energy
that is lost due to the condensation of $\phi$ is $\Delta V$ which
(from (\ref{con})) cancels $V_c$ exactly.

It is useful to see how this comes about in the Einstein frame,
where the action is
 \be\label{ei} S_{EH}+ S_{\rm inflation}= \int d^4 x \sqrt{-g}
  \left[ \frac12  R -3(\partial \phi)^2
   -\tilde{V}
 \right],\ee
with
 \be \tilde{V}= e^{2\phi^2} \left( V(\phi)+ V_{\rm ren} +\epsilon
 \psi\right).
% \left( V(\phi)+ V_{\rm ren} -\epsilon \psi \left) .
 \ee
Thus $m_{\rm eff}^2$ depends on the total vacuum energy $m_{\rm
eff}^2=-\gamma  + 4\left( V(\phi=0)+ V_{\rm ren} +\epsilon
\psi\right) $, which implies that an instability is developed at
$V=V_c=\gamma/4$ and that (since $\Delta \tilde{V}=\Delta V$) the
condition for the cancelation of the vacuum energy is indeed
(\ref{con}).

Two clarifications are in order. First, the cancelation of the
vacuum energy described above does not involve fine tuning $V_{\rm
ren}$. In fact the only assumption we made is that $V_{\rm
ren}+V(\phi=0)>V_c.$ Second, the feedback mechanism described above
depends only on $V(\phi)$ and not on $\epsilon$. Thus $\epsilon$ can
be taken to be small  so that on time scales of the age of the
universe the vacuum energy density does not change
significantly.\footnote{Non-perturbative effects such as
\cite{cdl,hm} impose a lower bound on $\epsilon$. Since these are
exponentially small this bound does not matter on time scales of the
age of the universe. Moreover, in multi-field generalizations of
this model one can avoid these effects.} This ensures that the
present value of the vacuum energy is of the order of its value at
the end of inflation.

The advantage of this model over \cite{ab} is that now, much like
in new inflation \cite{ni0,ni}, $\phi$ has  plenty of kinetic
energy that can  heat the universe. There are a couple of general
bounds on the reheating temperature, $T_r$. First, energy
conservation implies that $T_r^4 \leq V_c$ where we ignore factors
of order one and the number of fields. Second,  potential energy
is converted into kinetic energy only when the slow roll
approximation breaks down. This implies that $T_r^4 < \gamma$. In
our case $V_c\sim \gamma$ and so
 \be\label{tru} T_r \leq V_c^{1/4}  .\ee
To saturate this bound in the simplest scenario of reheating
 via  $\phi$
decay \cite{rh1,rh2} the decay rate should be $\Gamma_{\phi}\sim
m_{\phi}\sim \gamma^{1/2}$.

The discussion so far has not included quantum corrections to
$V(\phi)$ and as a result (\ref{tru}) does not imply  a relation
between $T_r$ and the present value of the vacuum energy density.
In what follows, we show how quantum effects and the present value
of the vacuum energy density impose an upper bound on $T_r$. For
concreteness we consider the following potential
 \be V(\phi)=-\frac12 m^2 \phi^2 + \frac14 g \phi^4.\ee
The condition for the cancelation of the vacuum energy, (\ref{con}),
in this case is
 \be\label{oa} m^2=g.\ee
Quantum corrections to (\ref{oa}) and more generally to the
effective potential, $V_{\rm eff}(\phi)$, yield a non-vanishing
vacuum energy at the end of inflation that we denote by $V_0$.
Technical naturalness of the model requires  that  $V_0 \sim
10^{-120}$. This is hard to achieve in non-supersymmetric theories
since the potential contains a relevant term that receives large
quantum corrections. In supersymmetric theories these corrections
are suppressed and (ignoring logarithmic factors) scale like
 \be V_0\sim g m^2 \sim m^4 .\ee
Thus technical naturalness of the vacuum energy implies that  $m
\sim 10^{-30}$ and that the upper bound on $T_r$ is approximately
$m^{1/2}\sim 10^{-15}\sim 1$ TeV.

Note that $m \sim 10^{-30}$ implies  a fifth force deviation from
GR at scales of the order of $1/m\sim 100$ microns since the
Newton constant depends on the expectation value of
$\phi$.\footnote{A different approach to the cosmological constant
problem that leads to a similar prediction is known as fat gravity
\cite{fg0,fg1,fg2}.} This prediction is particularly interesting
in light of current experiments \cite{ff,ff2} that should be able
to detect such a deviation  in the near future.

When supersymmetry is broken these corrections are enhanced. In
particular
 \be\label{m1} \delta m_{\phi}^2 \sim g \Delta m^2 ,~~~~~~\mbox{where}
 ~~~~~~~\Delta m^2=m_{\phi}^2-m_{\tilde{\phi}}^2.\ee
$\delta m_{\phi}^2$ represents  the quantum corrections to
$m_{\phi}^2$ and $m_{\tilde{\phi}}$ is the mass of the
super-partner of $\phi$. Technical naturalness implies  that
 \be\label{r} \delta m_{\phi}^2 \leq V_0 .\ee
Thus   $\Delta m^2$ is at most of the order of $10^{-60}$ and the
upper bound on the supersymmetry breaking scale in the $\phi$
sector is $M^{\phi}_{\rm SUSY   }\sim 10^{-30}$. Hence $\phi$
cannot be a field in the MSSM (or any other supersymmetric
generalization of the standard model) or in the hidden sector
responsible for the SUSY breaking. In fact such a low SUSY
breaking scale in the $\phi$ sector implies an upper bound on
$M_{\rm SUSY}$ at around $1$ TeV. This follows from the fact that
gravity always mediates SUSY breaking from one sector to the
other, which in our case gives
 \be M^{\phi}_{\rm SUSY   } \sim M_{\rm SUSY}^2~~~~~~
 \Rightarrow~~~~~~~M_{\rm SUSY}\sim \sqrt{M^{\phi}_{\rm SUSY   }}\sim 10^{-15}\sim
 1 \mbox{TeV}.\ee

Coupling with gravity is another source for quantum corrections to
the potential. It is easier to estimate these effects in the
Einstein frame (\ref{ei}). When $\tilde V=0$  there is a shift
symmetry, $\phi\to\phi+const.$ that prevents  gravity loops from
generating a potential for $\phi$. Hence corrections to $ \tilde{V}$
due to gravity loops are proportional to $ \tilde{V}$ and on
dimensional grounds the one loop contribution scales like $\delta
\tilde{V}\sim \tilde{V}\Lambda^2$, where $\Lambda$ is the cutoff
scale. In supergravity theories we expect the gravitino mass,
$m_{3/2}$, to play the role of $\Lambda$. Since $m_{3/2}\sim M_{\rm
SUSY}^2\sim m_{\phi}$ and $V_c\sim m_{\phi}^2$ we have
 \be \delta V_{\rm eff} \sim M_{\rm SUSY}^8,\ee
which leads to the same upper bound on  $M_{\rm SUSY}$  at around
$1$ Tev.

Alternatively, we may reach the same conclusion by applying
similar reasoning to the Kahler potential. Coupling with gravity
can generate corrections to the Kahler potential of the form
$|\phi|^{2n}$ with $n>1$. When $\tilde V=0$, the shift symmetry
implies that these corrections should vanish because otherwise
they would lead to terms of the form $|\partial \phi |^2
|\phi|^{2n-2}$ that do not respect the shift symmetry. Thus,
gravity   corrections to the Kahler potential scale like
$\tilde{V}$ and their effects on the effective potential scale
like $\tilde{V}^2\sim M_{\rm SUSY}^8$.\footnote{For example the
term $V(\phi)=m^2 |\phi|^2$ can generate a correction to the
Kahler potential that scales like $m^2 |\phi|^4$ which in turn
will generate $\delta V\sim (m^2 |\phi|^2)^2$.}

Another  quantum gravity effect  that should be considered is due
to higher order terms, such as $R^2$. These will modify the FRW
equations and as a result will shift $V_c$. This effect scales
like $V_c^2$ and since  $V_c\sim 10^{-60}$ it will not spoil the
naturalness of the vacuum energy.

The model described above relies heavily on low scale SUSY and it
is natural to ask whether there are other models that are
technically natural with high-scale SUSY or maybe even with no
SUSY at all. Below we describe a non supersymmetric model that
``almost'' does that and fails to be technically natural only
because of gravity loop effects.

Consider the following  action
 \be\label{uq}S_{EH}+ S_{\rm inflation}= \int d^4 x \sqrt{-g}
  \left[ \frac12 e^{- \phi^2} R-\frac12 (\partial\phi)^2+
   \frac{1}{16\pi^2}\left( \frac{\phi}{f} -\pi \right) \mbox{Tr}
 (F\wedge F)\right] ~~~ .\ee
The  shift by $\pi$ is needed for  $\phi=0$ to be  a maximum of
the potential induced by the instantons. If the gauge theory has a
weakly coupled fixed point (see e.g. \cite{bz}) then $V(\phi)$ is
dominated by the one-instanton contribution that takes the form
 \be\label{qa} V(\phi) = V_g \cos\left(\frac{\phi}{f}\right) , ~~~~~~\mbox{with}~~~~~~
 V_g=C e^{-8\pi^2/g^2}.\ee
$C$ is a constant that depends on the details of the UV cutoff on
the integration over the size of the instanton, which  we assume
to be  at the Planck scale.  Moreover, $C$ receives large quantum
corrections because quantum effects shift $1/g^2\to 1/g^2 +const.$
in the exponent. Thus without fine tuning only the order of
magnitude of $V_g$ can be fixed.

The nice feature of  this model is that $V_g$ drops out of the
condition for the cancelation of the vacuum energy, (\ref{con}),
that takes the form
 \be\label{con2} f^2=\frac18.\ee
The fact that $g$ is not exponentially small does not lead to
large corrections to (\ref{con2}) since as argued above loops
around the one instanton background can modify $V_g$ but not $f$.
The leading corrections to (\ref{con2}) come from two-instanton
effects  that scale like $V_g^2$. Thus naturalness of the vacuum
energy implies  that $V_g\sim10^{-60}$ and as before the upper
bound on $T_r$ is at the TeV scale.

Unfortunately, gravity loops spoil the naturalness of the model.
 A minimal coupling of an axion with gravity respects
the shift symmetry
 \be\label{sh}\phi\to\phi+2\pi  nf\ee
of (\ref{qa}) and thus can modify $V_g$ but not (\ref{con2}).
However, $\phi$ does not couple minimally with gravity. The
coupling $e^{-\phi^2}R$ breaks (\ref{sh}) and  generates terms in
the effective potential that do not respect (\ref{sh}) and
(without low scale SUSY) spoil the technical naturalness of the
model.

It is interesting  that in this model only  gravity loops are
problematic, while the cosmological constant problem is usually
associated with field theory loops. To be more precise, normally,
treating gravity semi-classically
 \be\label{sm}
  R_{\mu\nu}-\frac12 g_{\mu\nu} R=\langle
 T_{\mu\nu}\rangle, \ee
does not help much with the cosmological constant problem  since
field theory loops still  induce a large cosmological constant
term that affects the geometry via the semi-classical equation
(\ref{sm}). The amusing aspect of the axion model (\ref{uq}) is
that if one (wrongly) assumes (\ref{sm}) then a small vacuum
energy density is technically natural even without SUSY.

We of course would like to consider gravity at the  quantum level.
Hence  if we do not want to rely on low scale SUSY breaking then
we have to come up with an alternative way to suppress the
gravitational corrections to $V_{\rm eff}(\phi)$. One possible way
to do this is to replace the $R\phi^2$ trigger by a trigger that
depends on the Gauss-Bonnet combination
$R^2-4R_{\mu\nu}R^{\mu\nu}+R_{\mu\nu\sigma\rho}R^{\mu\nu\sigma\rho}$.
The topological nature of the Gauss-Bonnet term suppresses
corrections to $V_{\rm eff}(\phi)$ even without SUSY.\footnote{For
example the term $\lambda
(R^2-4R_{\mu\nu}R^{\mu\nu}+R_{\mu\nu\sigma\rho}R^{\mu\nu\sigma\rho})
\phi^2$ leads to $\delta m^2$ that scales only like $\lambda^2
\Lambda^{10}$.}  However, in the slow-roll approximation the
Gauss-Bonnet term scales like $V^2$ and not like $V$. This
complicates  the model and (as far as we can tell at the moment)
lowers  considerably the upper bound on $T_r$.

To summarize, the goal of this note was to suggest a simple
mechanism that renders the small value of the vacuum energy
density technically natural. The mechanism was illustrated  in the
context of a concrete example that  led to the following
predictions:

\nl (i) The upper bound on the reheating temperature is about $1$
TeV.

\nl (ii) SUSY is broken at around the TeV scale.

\nl (iii) There is a fifth force deviation from GR at scales of
the order of $100$ microns.

 We believe that it is reasonable to suspect that (ii) and (iii)
could be relaxed in generalizations of this model. It is, however,
hard to see how  models that use the same basic mechanism could
lead to $T_r$ much larger than $1$ TeV. As usual in models with
low scale inflation this makes the issue of baryogenesis subtle
because the electroweak scale is not too far below $1$ TeV. In our
case this issue is even more subtle since the interactions that
transfers the kinetic energy of $\phi$ into heat (and eventually
to baryons) should on the one hand be strong enough to yield
$T_r\sim 1$ TeV (or  at least to avoid the cosmological moduli
problem associated with a particle with such a small mass) and on
the other hand it should not spoil the technical naturalness of
the model. This implies that this interaction should involve only
derivatives of $\phi$ since interaction that involves $\phi$
itself and  gives $T_r\sim 1$ TeV will generate large quantum
corrections to $V_{\rm eff}(\phi)$.\footnote{There are also finite
temperature corrections to $V_{\rm eff}(\phi)$. These, however,
become negligible as the universe cools down and so they do not
modify the present value of the vacuum energy.} Even with
interactions that involve only derivatives of $\phi$ one needs to
make sure that subleading corrections to $V_{\rm eff}(\phi)$  are
suppressed. The symmetry (\ref{sh}) suggests that the axion model
is less sensitive to such corrections than  the $\phi^4$ model.

 \vspace{10mm}

\noindent {\bf Acknowledgements}

I thank C. Callan, L. McAllister, N. Seiberg  and P.  Steinhardt for
helpful discussions. I would also like to thank D. Chung for
pointing out a normalization typo. This material is based upon work
supported by the National Science Foundation under Grant No. PHY
0243680. Any opinions, findings, and conclusions or recommendations
expressed in this material are those of the author and do not
necessarily reflect the views of the National Science Foundation.

\end{document}